\documentclass[letterpaper, 10 pt, conference]{ieeeconf}  

\IEEEoverridecommandlockouts                              
\usepackage[pdftex]{graphicx}
\usepackage[cmex10]{amsmath}
\usepackage{amssymb}
\usepackage{import}
\usepackage{bm}

\usepackage{color}
\usepackage{units}
\usepackage{url}
\usepackage{float}
\usepackage{epstopdf}
\usepackage{stmaryrd}

\usepackage{units}
\usepackage{siunitx}

\usepackage{outlines}

\usepackage[font=small]{caption}
\usepackage{boldline}
\usepackage[flushleft]{threeparttable}

\usepackage[style=ieee,backend=biber,maxbibnames=1000]{biblatex}

\bibliography{main.bib}
\usepackage{mathtools}
\usepackage{balance}

\newcommand{\mnoshow}[1]{}

\newcommand{\figref}[1]{Fig.~\ref{#1}}
\newcommand{\secref}[1]{Section~\ref{#1}}

\title{\LARGE \bf
Safe Adaptive Cruise Control with Road Grade Preview and V2V Communication
}

\author{Roya Firoozi, Shima Nazari, Jacopo Guanetti, Ryan O'Gorman, Francesco Borrelli
\thanks{The authors are with the Department of Mechanical Engineering, at the University of California, Berkeley and University of Michigan. 
\newline  {\tt \{royafiroozi, jacopoguanetti, ryanogorman ,fborrelli \}@berkeley.edu}, \tt snazari@umich.edu}%
}

\begin{document}
\maketitle
\thispagestyle{empty}
\pagestyle{empty}
\begin{abstract}
We present the design of a safe Adaptive Cruise Control (ACC)  which uses  road grade and lead vehicle motion preview. The ACC controller is designed by using a Model Predictive Control (MPC) framework to optimize comfort, safety, energy-efficiency and speed tracking accuracy. Safety is achieved by computing a robust invariant terminal set.  The paper presents a novel approach to compute such set which is less conservative than existing  methods. The proposed controller ensures safe inter-vehicle spacing at all times despite changes in the road grade and uncertainty in the predicted motion of the lead vehicle. Simulation results compare the proposed controller with a controller that does not incorporate prior grade knowledge on two scenarios including car-following and autonomous intersection crossing. The results demonstrate the effectiveness of the proposed control algorithm.
\end{abstract}

\section{Introduction}
Recent advances in autonomous driving have accelerated the need for high performance and reliable Advanced Driving Assistance Systems (ADAS) which guarantee safety and comfort in various driving conditions. At the same time, connected car technologies begin to make their first appearance in the passenger vehicle market. Connected and automated vehicles will enhance mobility and safety by integrating autonomous driving with communication technologies. Connectivity enables ADAS to leverage Vehicle-to-Vehicle (V2V) and Vehicle-to-Infrastructure (V2I) communication to further improve performance \cite{GUANETTI201818}. 

Adaptive cruise control (ACC) is a widely used ADAS module that controls the vehicle longitudinal dynamics. ACC is triggered once a preceding vehicle is detected within a certain distance range from the ego vehicle. It automatically adjusts the vehicle acceleration and deceleration to maintain a proper minimum safe distance from the vehicles ahead. ACC enhances mobility, improves safety and comfort, and reduces energy consumption. The design of ACC based on Model Predictive Control (MPC) is common in the literature. For example, in \cite{Luo2010, article,4436187,7035071,7963297,8002647}, the authors present ACC design using MPC based on one or more particular performance criteria including traffic flow, energy efficiency and safety or comfort considerations.    

Conventional adaptive cruise control systems operate in two modes: ACC and Cruise Control (CC), depending on the presence or absence of the lead vehicle in detection range of the ego vehicle. In ACC mode the objective is to maintain a safe distance from the lead car, whereas in CC mode the objective is to track the reference velocity set by the driver or the maximum speed limit of the road. In the literature, ACC and CC modes are also referred to as distance tracking and velocity tracking modes, respectively. Discrete switching between the modes may result in aggressive control action or repetitive mode change. In \cite{SHAKOURI201112964, ZHENHAI2016581}, the authors suggest more sophisticated rule-based switching strategies to prevent chattering caused by switching. We formulate and design our proposed controller without employing any switch in order to adapt the velocity and distance automatically and smoothly, if a lead car is detected. 

In conventional ACC systems, the desired inter-vehicle gap can be specified as a constant space or time gap between the vehicles. The desired gap also might be defined via a constant time headway $t_{h}$ policy which relates the inter-vehicle desired distance $\Delta s_{des}$ to the current velocity $v$ of the ego vehicle as $\Delta s_{des} = \Delta s + t_{h}v$, 
where $\Delta s$ is the static gap or the distance between the cars when they are at standstill and $t_{h}v$ is the dynamic gap which changes with velocity. In general these simple approaches do not guarantee safety. To guarantee safety at all times, the theory of invariant sets can be used to compute the safe distance. In \cite{ALAM201433}, reachability analysis with level set method is proposed to compute the safe set for heavy duty platooning. The authors of \cite{7336573} use a kinematic model to describe the system as a linear time-invariant (LTI) system and apply backward reachable set analysis to calculate a polytopic control invariant set for ACC. In  \cite{turri2017cooperative}, a control invariant set is presented as the safe set for platooning by considering the lower and upper bounds on the acceleration. The aforementioned studies compute the control invariant set by bounding the unknown parameters such as road slope or the front car's motion. However, computing the control invariant safe set by considering the lower and upper bounds on the unknown parameters can lead to an overly conservative safe set and thus an undesirable large gap between the vehicles. In the proposed study, instead of bounding these parameters, we assume the availability of road grade preview from a high fidelity map and the availability of lead car's future state trajectory from a V2V communication device. Employing this parameters for computing the control invariant safe set, instead of their boundary values, yields a less conservative safe set and consequently a closer inter-vehicle safe distance. 

Road grade can considerably affect the ACC controller performance. The work in \cite{8002647} considers an ACC which takes into account road elevation data to improve energy efficiency. In \cite{javad} real-time estimation of road grade and vehicle mass are utilized to improve comfort. In this study, we investigate how the ACC controller performance improves in terms of safety, comfort, tracking accuracy and energy efficiency, by exploiting the road grade knowledge. 

The contribution of this paper can be summarized as follows:
\begin{itemize}
\item We propose the design of an ACC controller which incorporates the road grade preview information using a prior grade map of the road to predict the vehicles' longitudinal dynamics. Furthermore, the proposed controller is formulated to switch automatically and smoothly between distance and velocity tracking.  
\item A numerical approach to compute a control invariant set which makes use of the road grade preview and the future state trajectory of the leading car, transmitted through V2V communication, is presented. This set is incorporated as terminal constraint of the ACC controller to guarantee safety.    
\item We demonstrate through two example scenarios that the proposed approach is more efficient compared to the controller that does not use grade preview. 
\end{itemize}

The rest of the paper is structured as follows: \secref{secPreliminaries} introduces some preliminaries about vehicle model and road grade map generation. \secref{secProblemDef} describes the problem statement and the MPC formulation. \secref{secSafeSet} explains the design of control invariant set. \secref{secScenarios} describes car-following and autonomous intersection passing scenarios as two example applications of the proposed approach. \secref{secResults} illustrates the simulation results and finally \secref{secConclusion} makes the concluding remarks. 

\section{Preliminaries}
\label{secPreliminaries}

\subsection{Vehicle Model}
\begin{figure}
\centering
    \includegraphics[]{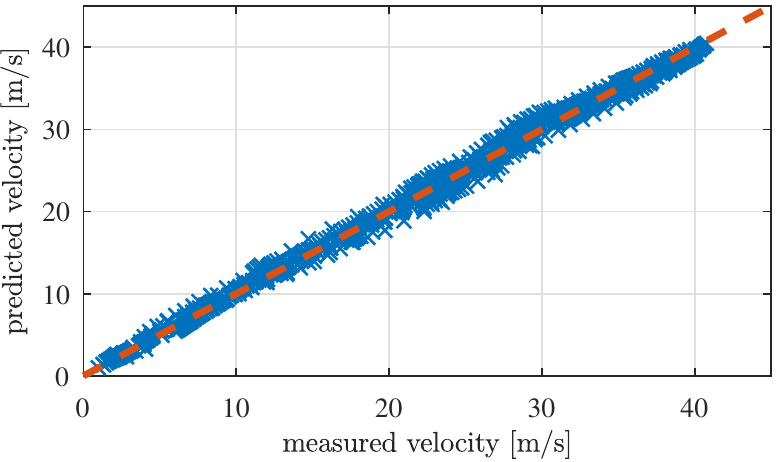}
\caption[]{Fit of the identified longitudinal dynamics model to the experimental data.}
\label{fig:model_fit}%
\end{figure}

The vehicle is modeled as a point mass moving along a road. The system state at time t is
\begin{equation*}
    x(t) = [s(t) \ v(t)]^{T},
\end{equation*}
where $s(t)$ and $v(t)$ are the vehicle position and velocity, respectively.
The longitudinal motion of the vehicle can be described by the following equations
\begin{equation}\label{eq:vehicle_model}
    \begin{aligned}
        \dot{s} &= v\\
        \dot{v}&= \frac{1}{m} (\underbrace{F_{t}-F_{b}-F_{a}-F_{r}-F_{g}}_{\text F_{total}}),
    \end{aligned}
\end{equation}
where $F_{t}$ and $F_{b}$ are traction and braking forces, respectively and $F_{total}$ is the total longitudinal force. The aerodynamic drag is determined by vehicle speed $v$, air density $\rho$, air drag coefficient $C_d$ and frontal area $A_f$.
\begin{equation}\label{eq:drag_force}
    F_{a} = \frac{1}{2}\rho C_{d}A_{f} v^2.
\end{equation}
The rolling resistance is defined as 
\begin{equation}\label{eq:rolling_force}
    F_{r} = m g C_r \cos \theta,
\end{equation}
where $g$ is gravitational force and $C_{r}$ is rolling friction coefficient. 
The gravity force due to road slope can be expressed as
\begin{equation}\label{eq:gravity_force}
    F_{g} = m g \sin\theta.
\end{equation}
The system \eqref{eq:vehicle_model} is discretized using Euler method with constant sampling time interval $\Delta t$.
\begin{equation}\label{eq:discrete_system}
    \begin{aligned}
        s(t+1) &= s(t) + v(t) \Delta t,\\
        v(t+1) &= v(t) + \frac{\Delta t}{m} F_{total}\\
    \end{aligned}
\end{equation}
The parameters of model~\eqref{eq:discrete_system} are then estimated by
nonlinear least squares from data collected on a test vehicle, at various speeds and accelerations.
All the data are collected in a proving ground on level roads with homogeneous surfaces, i.e. $\theta = 0$ and $C_r$ is constant for the purpose of model identification.
Fig.~\ref{fig:model_fit} shows the fit of the identified model to the experimental data, while Table~\ref{tab:parameters_fit} summarizes the identified model parameters.
Specifically, we use 7 datasets for a total duration of about 16 minutes; the predicted data in Fig.~\ref{fig:model_fit} are generated simulating model~\eqref{eq:discrete_system} with the identified parameters and the measured input torque from the initial condition to the end of the dataset.

\begin{table}
\centering
\caption{Vehicle Model Parameters}
\begin{tabular}{ c  c  c  c }
\hline
$m$         & vehicle mass              & kg        & 2278\\
$A_f$       & vehicle frontal surface   & m$^2$     & 2.63\\
$\rho$      & air density               & kg/m$^3$  & 1.206\\
$C_d$       & vehicle drag coefficient  & -         & 0.2791\\
$C_r$       & vehicle roll coefficient  & -         & 0.0089\\
$\Delta t$  & sampling time             & s         & 0.5\\
\hline
\end{tabular}
\label{tab:parameters_fit}
\end{table}

\subsection{Road Grade Map Generation}
Road topology data is a valuable source of information in autonomous driving applications. On roads with up-down slopes, ADAS systems like ACC benefit from accurate prior-known road grade. ACC can exploit the grade preview extracted from a map to ensure keeping a safe distance from the lead vehicle. Different methods for generating a high-resolution grade map of the road have been discussed in our previous work \cite{04167}. The road grade profile can be obtained using a survey vehicle equipped with high-precision GPS system. An alternative approach is using APIs like Google Elevation API. Google Elevation API provides elevation data for all locations on earth surface. Elevation data can be queried for specific coordinates to generate road elevation profile. In this study, we utilize Google Elevation API to generate the road grade profile. 
\section{Problem definition and formulation}
\label{secProblemDef}
The goal is to design an ACC and CC controller with safety guarantees.
The proposed ACC/CC controller is formulated as Model Predictive Control (MPC) which repeatedly solves the following constrained finite horizon nonlinear optimization problem:
\begin{subequations}\label{eq:mpc_formulation}
\begin{align}
    & \underset{{U_{t\rightarrow t+N|t}}}{\text{minimize}} 
    \label{eq:tracking}
    & &J = Q\sum_{k=t}^{t+N-1}||v(k|t)-v_{\text{ref}}||_{2}\\
    \label{eq:energy_efficiency}
    & & & + R_{u}\sum_{k=t}^{t+N-1} ||u(k|t)||_{2}\\
    \label{eq:jerk}
    & & &+ R_{\Delta u} \sum_{k=t}^{t+N-1}||\Delta u(k|t)||_{2}\\
    \label{eq:terminal_cost}
    & & &+ P||v(t+N|t)-v_{\text{ref}}||_{2}\\
    &\nonumber \textrm{subject to}\\ 
    \label{eq:dynamic_const}
    & & & x(k+1|t) = f(x(k|t),u(k|t),\theta(k|t)),\\
    \label{eq:state_bound}
    & & & v_{\min} \leq v(k|t) \leq v_{\max},\\
    \label{eq:input_bound}
    & & & u_{\min} \leq u(k|t) \leq u_{\max},\\
    & & & x(t|t) = x(t),\\
    & & & \nonumber d_{\text{safe}}\big(s_{\text{lead}}(k|t)-s(k|t),v(k|t)\\
    & & & v_{\text{lead}}(k|t), (\theta(t|t),...,\theta(t_{\text{stop}}|t)\big)\\
    \label{eq:safety_constraint}
    & & & \quad \leq s_{\text{lead}}(k|t) - s(k|t),\\
    & & & \nonumber k = t,...,t+N,
\end{align}
\end{subequations}
where $x(k|t)=[s(k|t) \quad v(k|t)]^T$ and $u(k|t)$ are the state and control input at step $k$ predicted at time $t$, respectively.
The MPC horizon is $N$ and $U_{t\rightarrow t+N|t}$ denotes the sequence of control inputs $\{u(t|t),...,u(t+N-1|t)\}$.
The multi-objective quadratic cost function $J$ represents the trade-off between minimizing reference tracking error \eqref{eq:tracking}, control effort \eqref{eq:energy_efficiency} and jerk \eqref{eq:jerk} and $Q$, $R_{u}$ and $R_{\Delta u}$ are their corresponding weight factors, respectively.
The terminal cost \eqref{eq:terminal_cost} is weighted by $P$. The state and input are constrained as \eqref{eq:state_bound} and \eqref{eq:input_bound} to lie within their lower and upper bounds denoted as $v_\text{{min}}$, $u_\text{{min}}$ and $v_\text{{max}}$, $u_\text{{max}}$, respectively.
The reference velocity $v_\text{ref}$ is the desired velocity set for the ego vehicle by cruise control.
The vehicle longitudinal dynamics \eqref{eq:discrete_system}, denoted as $f$, introduce nonlinear constraints, parameterized by the road grade $\theta$.
Since the road grade map is available as position-dependent data, the controller has to convert it from spatial domain to time domain to make use of the grade preview.
To do so, the ego car's velocity is assumed to remain constant over the MPC horizon and grade data corresponding to the predicted positions is extracted.
The lead car's velocity and position at step $k$ predicted at time $t$ are denoted as $v_{\text{lead}}(k|t)$ and $s_{\text{lead}}(k|t)$, respectively. The predicted future trajectory of the lead car is transmitted through V2V communication. 
The safe distance between the two vehicles, denoted as $d_{\text{safe}}$, is obtained by calculating the control invariant set, described in \secref{secSafeSet}. $d_{\text{safe}}$ is a function of the position and velocity of the ego and lead vehicles, and of the road grade from the current time $t$ till $t_{\text{stop}}$, which is the time at which the lead car comes to full stop if it exert its maximum braking force. 
The optimal solution of the problem \eqref{eq:mpc_formulation} is 
\begin{equation*}
U^*(t) = \{u^*(t|t),...,u^*(t+N-1|t)\},
\end{equation*}
and the receding horizon control law is obtained by applying the first control input
\begin{equation}\label{optimal_policy}
u_{\textrm{MPC}}(t) = u^*(t|t). 
\end{equation}
The above formulation includes switching behavior from distance tracking (ACC) mode to reference velocity tracking (CC) mode implicitly. In other words, there is no explicit term for discrete switching between the two modes, but the controller is designed to operate in ACC mode as long as the front vehicle is detected and automatically and smoothly adapts the velocity to CC reference once there is no vehicle in the front.

The multi-objective cost $J$ in problem \eqref{eq:mpc_formulation} simultaneously fulfills multiple performance criteria. To assess the performance of the controller we introduce a performance index for each objective. The first term of the cost function \eqref{eq:tracking} denotes reference tracking error and \textit{tracking performance index} can be defined as 
\begin{equation}\label{tracking_index}
    \text{Tracking Performance Index} = \sum_{t=0}^{T}|v(t)-v_{\text{ref}}|,
\end{equation}
where $T$ is the total time of simulation or experiment, $v(t)$ is the ego vehicle velocity at time $t$ obtained as the closed-loop state of system  \eqref{eq:discrete_system} controlled with $u_{\textrm{MPC}}$ described in \eqref{optimal_policy}.   
The second term of the cost \eqref{eq:energy_efficiency} penalizes the control effort and presents energy consumption criteria. We introduce the \textit{energy consumption performance index} as
\begin{equation}\label{energy_index}
    \text{\small Energy Performance Index} = \sum_{t=0}^{T}\text{max}(0,u_{\textrm{MPC}}(t)),
\end{equation}
assuming no penalty for braking. 
The third term in the cost \eqref{eq:jerk} minimizes the change of acceleration (jerk) which is a performance criteria for longitudinal ride comfort. The \textit{comfort performance index} is 
\begin{equation}\label{jerk_index}
    \text{\small Comfort Performance Index} = \sum_{t=0}^{T}|u_{\textrm{MPC}}(t+1)-u_{\textrm{MPC}}(t)|.
\end{equation}
In both \eqref{energy_index} and \eqref{jerk_index}, $u_{\textrm{MPC}}(t)$ is the closed-loop control input described in \eqref{optimal_policy}.

\section{Control Invariant Set}
\label{secSafeSet}
The adaptive cruise control is considered to be safe if the follower vehicle can avoid collisions with the front vehicle regardless of the front vehicle action.
Consider the combined model of the front and follower vehicles in Adaptive Cruise Control (ACC) problem,
\begin{align}
    \label{eq:3statesEOM}
    \left[
    \begin{array}{c}
    \dot{s}_{l}-\dot{s}_{e}\\
    \dot{v}_e\\
    \dot{v}_l
    \end{array}
    \right]
    =
    \left[
    \begin{array}{c}
    v_l - v_e\\
    \frac{1}{m_e}\big( F_{t,e} - F_{b,e} - F_{a,e}-F_{r,e}-F_{g,e}\big)\\
    \frac{1}{m_l} F_{total,lead}\\
    \end{array}
    \right]
\end{align}
$F_{t,e}$, $F_{b,e}$, $F_{a,e}$, $F_{r,e}$ and $F_{g,e}$ denote the ego vehicle tractive force, braking force, aerodynamic drag force, rolling resistance force, and gravity force respectively defined in \eqref{eq:drag_force}-\eqref{eq:gravity_force}. $m_e$ and $m_l$ are the ego and lead vehicle masses, respectively.
$v_e$ and $s_e$ are the ego vehicle velocity and position.
$v_l$ and $s_l$ are the lead vehicle velocity and position.
In this system, the ego (follower) vehicle braking/tractive force is the control input and the lead (front) vehicle's total longitudinal force denoted as $F_{total,Lead}$ is treated as disturbance to the system \eqref{eq:3statesEOM} and described based on \eqref{eq:vehicle_model}. The set of all admissible states for this system are defined as follows:
\begin{align*}
\mathcal{X} = \{[s_l-s_e \hspace{3pt} v_e \hspace{3pt} v_l \hspace{3pt} ]^T: \hspace{3pt} (s_l-s_e) >l_{min}, \hspace{3pt} v_e>0, \hspace{3pt} v_l>0 \}
\end{align*}
where $l_{min}$ is the minimum required distance between the mass center of the two vehicles.
We define the robust control invariant set $\mathcal{C} \subseteq\mathcal{X}$ as 
a set with the following property:
\begin{equation*}
\begin{aligned} & \textrm{if} \ x(t) \in \mathcal{C} 
\Longrightarrow \exists u(t) \in \mathcal{U} \ \textrm{such that} \\ 
& f(x(t),u(t),\theta(t)) \in \mathcal{C}, \\
&\forall  F_{total,Lead}(t) \in \mathcal{F}_{total,Lead}, \
, \ \forall t \in \mathbb{N}^+.
\end{aligned}
\end{equation*}
The set $\mathcal{C}$ is a function of the road slope $\theta(\cdot)$ at time $t$ until the time $t_{stop}$ at which the front car comes to full stop after exerting the brake at time $t$. Therefore, $\mathcal{C}=\mathcal{C}(\theta(t),...,\theta(t+t_{stop}))$. $x(t)$ denotes the state of the system \eqref{eq:3statesEOM} at time $t$,  $\mathcal{U}$ is the input feasible set \eqref{eq:input_bound}, $f$ represents the system dynamics \eqref{eq:3statesEOM}.
$\mathcal{F}_{total,Lead}$ is the set of all possible longitudinal forces of the lead vehicle.
Hence, the robust control invariant set $\mathcal{C}$ for the above ACC system is such that for any maneuver of the front vehicle, there is a control signal that keeps the system \eqref{eq:3statesEOM} within $\mathcal{C}$ for all future times \cite{borrelli2017predictive}.

A conservative estimation of the closed form of $\mathcal{C}$ for this problem is presented in \cite{turri2017cooperative} that computes the control invariant set by finding the bounds on the road grade and calculating the bounds on accelerations of the lead car.
In this work a numerical method is employed to compute the boundary of the safe set more accurately using the road slope preview and future trajectory of the lead car.
By using such preview information, we calculate the exact maneuver of the front car by assuming it exerts maximum braking force at each time step, using a two-step approach.
\begin{itemize}
\item In the first step, the equations of motion of the front vehicle are integrated using forward Euler discretization, assuming the front vehicle applies its maximum braking force, $F_{b,l}^{\max}$,
\begin{subequations}
\begin{align}
\label{eq:FEuler_V1V}
\nonumber v_l(k+1) &= v_l(k)+\frac{dt}{m_l}\Bigg( - F_{b,l} ^{\max}\\
&\nonumber -\frac{1}{2}\rho C_{d,l} A_{f,l} v(k)^2 - m_l g C_{r,l}\cos\Big(\theta(s_l(k)\Big),\\
& -m_lg\sin\Big(\theta \big(s_l(k)\big)\Big)\Bigg),\\
\label{eq:FEuler_V1S}
s_l(k+1) &= s_l(k)+dt v_l(k)
\end{align}
\end{subequations}
where the initial condition for \eqref{eq:FEuler_V1V} and \eqref{eq:FEuler_V1S} is the current velocity and position of the front vehicle.
In \eqref{eq:FEuler_V1V} the road grade depends on the position of the front vehicle at each time instant. $k$ is the integration interval.
The integration stops when the front vehicle reaches zero velocity, $v_l(k) = 0$.
Define the position when the front vehicle reaches $v_{l} = 0$ as $ s_l^{stop}=s_l(k)$. 
\item In the second step, the equations of motion of the following vehicle are integrated backwards in time, assuming that the following vehicle is also applying its maximum braking force, $F_{b,e}^{max}$, with the initial condition of $v_e(0) = 0$ and $s_e(0) = s_l^{stop}-l_{min}$.
The integration continues till the velocity reaches its upper bound, $v_e(k) = v_{max}$. 
\begin{subequations}
\begin{align}
    \label{eq:FEuler_V2V} 
    v_e(k-1) &= \nonumber v_e(k)-\frac{dt}{m_e}\Bigg( - F_{b,e} ^{\max}\\ &-\nonumber \frac{1}{2}\rho C_{d,e} A_{f,e} v_e(k)^2-m_e g C_{r,e}\cos\Big(\theta(s_e(k)\Big)\\ &-m_eg\sin\Big(\theta \big(s_e(k)\big)\Big)\Bigg)\\\label{eq:FEuler_V2S}
    s_e(k-1) &= s_e(k)-dt v_e(k)
\end{align}
\end{subequations}
\end{itemize}
This method, although computationally more expensive compared to the closed form, takes into account the exact road grade profile. The minimum safe distance at each step time of the above integration is computed as follows,
\begin{align}
    d^{min}_{safe}(k) = s_l(0) - s_e(k)
\end{align}
Note \eqref{eq:FEuler_V1S} and \eqref{eq:FEuler_V2S} require the backward and forward simulation of absolute vehicle positions. One can rewrite the dynamics to highlight that only $\Delta s = s_l - s_e$ is important, as expressed in \eqref{eq:3statesEOM}. 

\begin{figure}[t]%
\centering
\includegraphics[trim= 0cm 0cm 0cm 0cm, clip=true, width=0.35\textwidth]{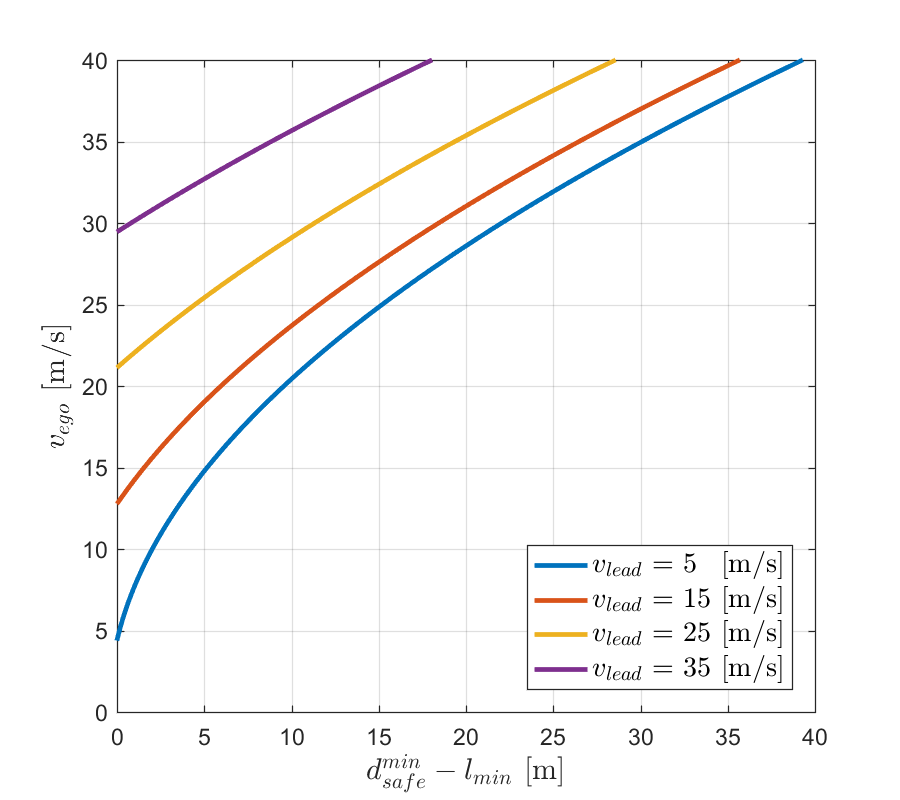} 
\caption[]{The boundary of control invariant set for the ACC problem.}
\label{fig:safe_set}%
\end{figure}

\begin{figure}[t]
    \centering
    \includegraphics[trim= 0cm 0cm 0cm 0cm, clip=true, width=0.35\textwidth]{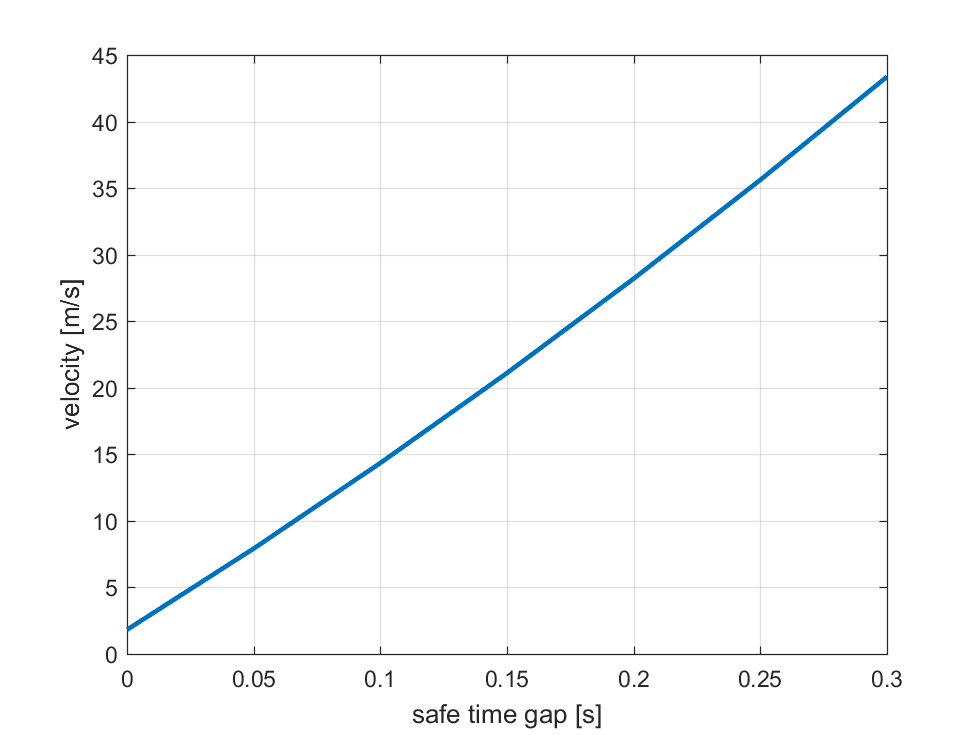} 
    \caption{Safe time gap corresponding to the safe minimum required distances reported at \figref{fig:safe_set}. }
    \label{fig:time_gap}
\end{figure}

\figref{fig:safe_set} shows examples of the safe set boundary for various velocity values of the front car.
The safe set boundary is obtained by fitting a second-degree polynomial on data points calculated using the above two-step approach.
Sufficiently small integration interval increases the data points and results in achieving higher fitting accuracy.
The polynomial as the function of ego vehicle velocity defines the required minimum safe distance which is imposed as safety constraint \eqref{eq:safety_constraint} in the MPC formulation \eqref{eq:mpc_formulation}. Also \figref{fig:time_gap} demonstrates the safe time gap plot corresponding to the \figref{fig:safe_set} safe distances. The time gap is computed for the cases that ego and lead vehicles' velocity are the same.  

To guarantee feasibility of problem \eqref{eq:mpc_formulation} at all times, persistent feasibility should be proven by showing the existence of a feasible control sequence at all times when starting from a feasible initial point.
Assume that at time $t_{0}$ a lead vehicle be in front of the follower vehicle and the problem \eqref{eq:mpc_formulation} be feasible.
Let $x(t)$ be the state of the system \eqref{eq:vehicle_model} in closed-loop with the MPC controller \eqref{eq:mpc_formulation} at $t>t_{0}$.
Since the problem is feasible at $x(0)$, there exist an optimal control sequence $\{u_0^*, u_1^*, ... , u_{N-1}^*\}$ at $t_0$.
Apply $u_0^*$ and let the system evolve to $x(1)$.
At $x(1)$, apply $u_{min}$ at the end of the MPC horizon.
The control sequence $\{u_1^*, u_2^*, ... , u_{min}\}$, since $u_{min}$  is input feasible and state feasible.
In fact, from the construction of the safe set, by applying $u_{min}$ at step $N$ will guarantee that $x(N+1)$ will be at a safe distance $d_{\textrm{safe}}(N+1)$.
In conclusion, the closed-loop system is persistently feasible. 

\section{Example Application Scenarios}
\label{secScenarios}
\begin{figure}[t]
\centering
\includegraphics[width=\linewidth]{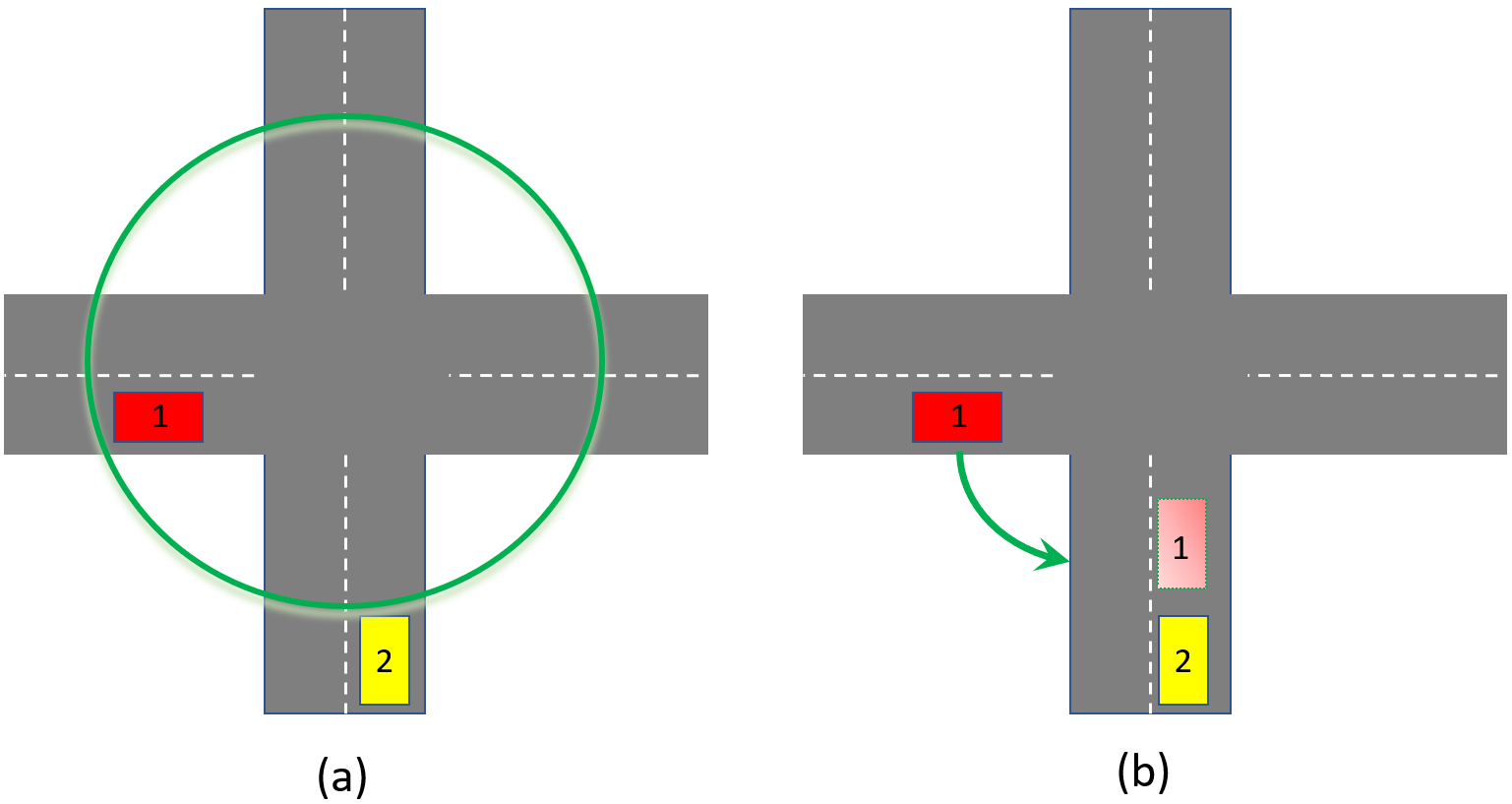}
\caption{a) The radius of green circle around the intersection represents the range of V2V communication. Since the red car has entered the circle sooner than the yellow one, the red car is prioritized to pass the intersection. So the red car is the leader and the yellow car is the follower. The yellow car has to adapt its speed according to the leader car speed and pass through the intersection with second priority.
b) The lead car (red) is projected at each time step in front of the follower car (yellow) which has the second priority. Light red rectangle shows the virtual car in front of the yellow car.}
\label{fig:communication_circle}
\end{figure}

The application of an ACC controller that adapts the vehicle's longitudinal velocity based on the other vehicle's states, communicated over V2V network, is not restricted to the car-following scenario. The same formulation of ACC can be extended further to other applications like autonomous intersection crossing. In an uncontrolled intersection, two cars that simultaneously approach an intersection at crossing directions, communicate with each other and adapt their speed to avoid collision and pass the intersection safely and efficiently \cite{Rios-Torres2017a}. 
A possible approach to coordinate the vehicles in an autonomous intersection is to define a circle centered at the center of the intersection and with radius of the range of communication $R$, as shown in \figref{fig:communication_circle}(a). Once the cars enter this virtual circle, they communicate with each other and assign the priority to each other based on their current velocity. Priority assignment for autonomous intersection is not the focus of this work and has been discussed in \cite{Rios-Torres2017a}. We assume we already know which car is prioritized to pass through the intersection first and play the role of the leader for the other car. The car with lower priority is the one that runs the ACC controller and adapts its velocity and distance to the intersection corresponding to the lead car. At each time step we project the lead car in the crossing direction in front of the ego car, as shown in \figref{fig:communication_circle}(b), as a virtual car that the ego vehicle follows.

\section{Simulation Results and Discussion}
\label{secResults}

We modeled and solved the optimization problem \eqref{eq:mpc_formulation} using YALMIP \cite{YALMIP} and IPOPT. The vehicle model and MPC parameters are presented in Table \ref{tab:parameters_fit} and \ref{tab:controller_parameters}, respectively.
To conduct realistic simulations, we drove a car on roads with significant slope (located near UC Berkeley) and collected latitude, longitude and velocity data.
We generated the grade maps of the roads by querying the Google Elevation API, and obtaining elevation data for the collected latitudes and longitudes.
The recorded velocity data of the car is considered as the front car's velocity in the simulations.
We ran simulation tests with different real roads grade profiles and assessed the performance of the proposed approach against the baseline approach.
Our proposed ACC considers the road grade preview in both planning and safe set calculation, while the baseline controller does not include knowledge of grade. 

\begin{table}[h]
\centering
\caption{MPC Controller Parameters}
\begin{tabular}{ c  c  c  c }
\hline
$v_{\min}$ & minimum velocity & m/s & 0\\
$v_{\max}$ & maximum velocity & m/s & 30\\
$u_{\min}$ & minimum control input & kN & -3 \\
$u_{\max}$ & maximum control input & kN & 3 \\
$\Delta t$ & sampling time & s & 0.2\\
$Q$ & tracking error cost weight & - & 10 \\
$R$ & control effort cost weight & - & 1 \\
$R_{\Delta u}$ & jerk cost weight & - & 10 \\
$P$ & terminal cost weight & - & 100 \\
\hline
\end{tabular}
\label{tab:controller_parameters}
\end{table}

\begin{figure}[t]
\centering
\includegraphics[width = \linewidth]{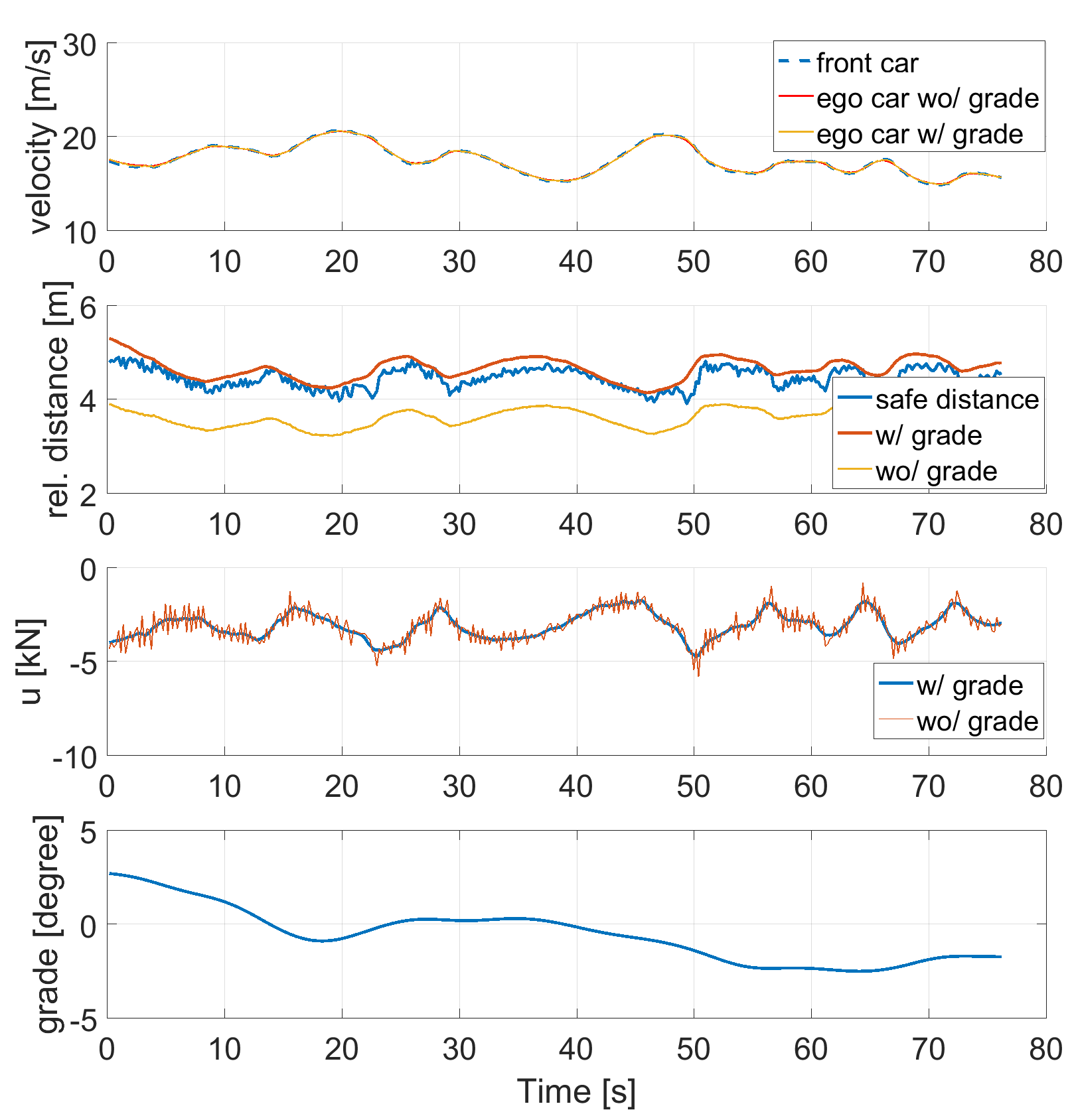}
\caption{Car-following scenario: w/grade is the case that grade knowledge is available to the controller (proposed controller), wo/grade is the case that grade knowledge is not available (baseline controller). The front car's velocity and road grade profile are realistic data.}
\label{fig:car_following_results}
\end{figure}

\figref{fig:car_following_results} represents the results for a car-following scenario.
The first graph compares the velocity of the front and the ego car; the tracking accuracy with and without knowledge of grade is essentially the same.
The second graph depicts the relative distance between the ego and front vehicles, with and without grade knowledge.
The safe distance in the graph represents the minimum required distance between the vehicles calculated by the robust control invariant set.
The relative distance obtained by the proposed controller 
is lower-bounded by the safe distance.
However, the relative distance obtained by the baseline controller (that has no knowledge of the road grade) is not safe, since the relative distance violates the minimum required safe distance.
The third graph represents the control input $u_{\text{MPC}}$.
As seen, the control input obtained by the proposed controller is much smoother compared to the baseline controller.
The fourth graph depicts the road grade profile obtained by Google Elevation API. 

To assess the performance of both controllers quantitatively, we used the performance indexes introduced in \secref{secProblemDef}. 
Table \ref{tab:result_table} compares, for various simulation runs, the energy consumption, tracking and comfort performance indexes when the road grade preview is either known or not known.
The total cost, which is the sum of all performance indexes, is also reported; the average total cost without grade knowledge is considerably higher than with grade preview.
 
\begin{table*}[t]
\centering
\begin{tabular}{|c|c|c|c|c|c|c|c|c|}
\hline
Road Segment &\multicolumn{2}{c|}{Energy Consumption}& \multicolumn{2}{c|}{Tracking} & \multicolumn{2}{c|}{Comfort} & \multicolumn{2}{c|}{Total Cost}\\
\hline \hline
 & W/ grade & W/O grade  & W/ grade & W/O grade & W/ grade & W/O grade & W/ grade & W/O grade  
\\
\hline
1& 311.9 & 324.3 & 68.5 & 127.2 & 49.0 & 50.0 & 430.3 & 501.5\\
2& 615.1 & 625.2 & 47.1 & 64.2  & 60.9 & 86.5 & 723.1 & 775.9\\
3& 0 & 0.81 & 28.7 & 30.9 & 30.8 & 189.7 & 59.6 & 221.4\\
4& 0 & 0 & 45.0 & 37.0 & 52.3 & 205.6 & 97.3 & 242.6\\
5& 0 & 0.11 & 32.6 & 30.6 & 36.2 & 286.4 & 68.8 & 317.1\\
6& 149.1 & 158.0 & 70.2 & 106.0 & 109.8 & 175.0 & 329.1 & 439\\
\hline
\end{tabular}
\caption{Performance indexes for different segments of a hilly road located near UC Berkeley are reported for two cases of w/grade (proposed controller with grade knowledge) and wo/grade (the baseline controller without grade knowledge). The average of the total costs over all the segments is  416.3 W/O grade knowledge which is considerably larger then the 284.7 obtained with the proposed controller.}
\label{tab:result_table}
\end{table*}

For the autonomous intersection scenario the results are similar to the car-following case, and are not reported here for brevity.
We just note that in this scenario, the road grade profiles for the leader and follower cars are different; at each time step the vehicles' distances to the center of intersection are measured and their relative distance is calculated by projecting the lead car in front of the follower car.

Both scenarios (car-following and autonomous intersection crossing) are set up in the PreScan simulation environment.
PreScan has an interface with MATLAB/Simulink and is a suitable platform for developing ADAS systems as well as modeling V2V communication.
We also modeled the road grade for both scenarios in PreScan environment with sinusoidally varying profiles.
The lead vehicle velocity is also generated as another sinusoidal profile.
The PreScan video for both scenarios is available online.\footnote{\url{https://www.youtube.com/watch?v=Qi9Lehtvqjc}}
In the car-following video the lead car is moving with a sinusoidal velocity profile, until it applies full braking and comes to full stop in the middle of the road.
The follower car is able to maintain the safe distance and avoid collision using the proposed approach.
This is a visualization that shows how the control algorithm is robust with respect to aggressive maneuvers of the lead car on roads with any arbitrary grade profile.
In the autonomous intersection video the vehicles communicate with each other and after prioritization, the follower adapts its velocity based on the lead car's velocity and avoids collision by maintaining the safe distance calculated using the control invariant set.
Both vehicles pass the intersection safely for an arbitrary road grade profile using the proposed approach.  

\begin{figure}[t]
    \centering
    \includegraphics[width = \linewidth]{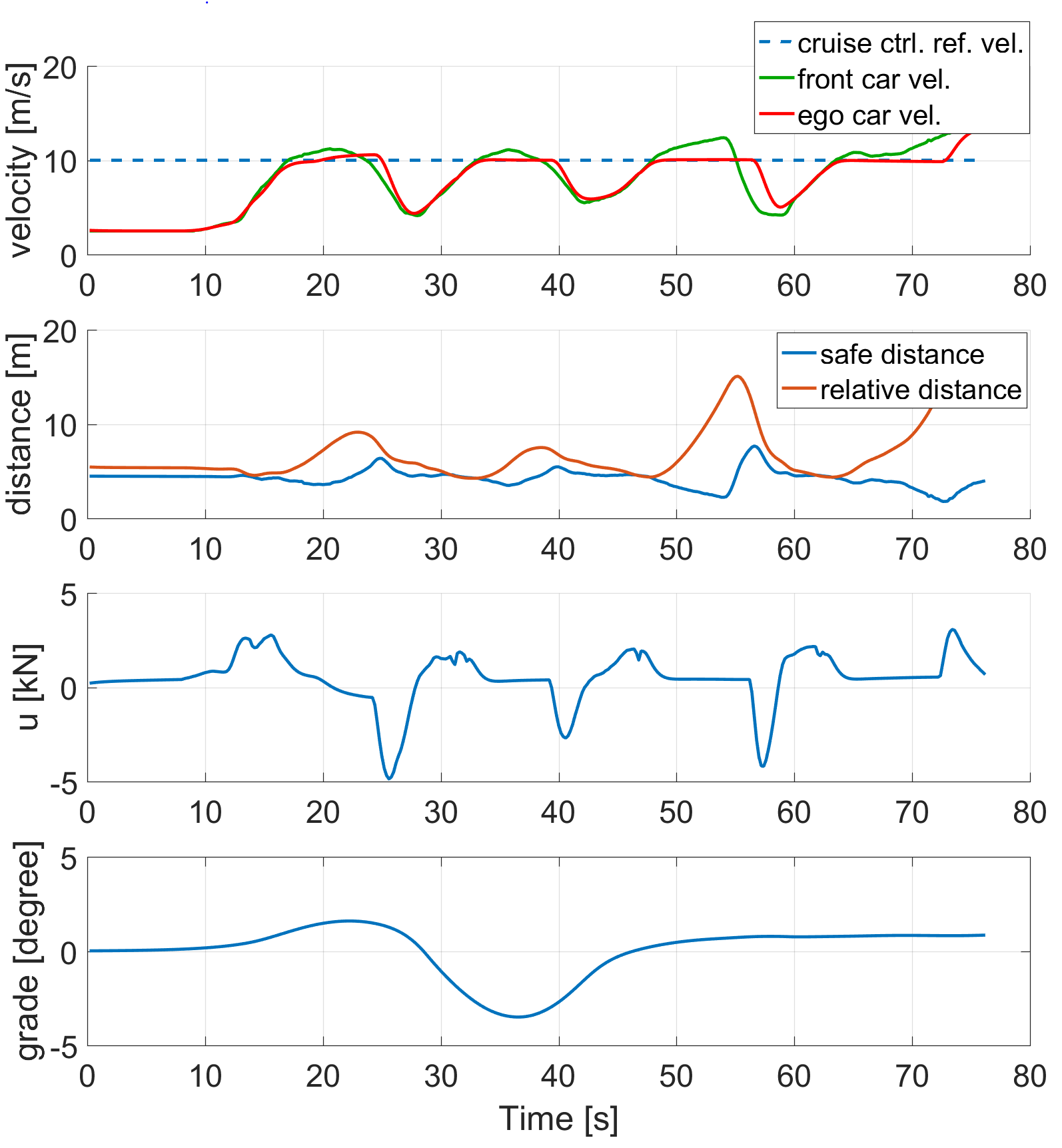}
    \caption{Automatic, smooth and safe switching between ACC and CC modes.}
    \label{fig:switching_results}
\end{figure}

As previously described, the proposed controller is capable of automatically, safely and smoothly switching between CC and ACC modes.
\figref{fig:switching_results} shows the results for this case.
The front car's velocity is obtained by collecting real driving data and the associated road grade map is created using Google Elevation API.
The cruise control speed is set to a constant value.
By comparing the velocity and distance plots, we can see that the ego car velocity tracks the cruise control reference velocity when the front car is far, for example between the time of $15$ to $25$.
Afterwards, for example between the time $28$ to $34$, since the front car's velocity is less than the cruise control reference speed, and the distance between the cars is closer, the ego car tracks the front cars' velocity instead.
The relative distance shown in the second plot is also lower-bounded by the safe distance.  

\section{Conclusion}
\label{secConclusion}
We designed a performance-enhanced ACC controller that exploits preview information about road grade as well as lead vehicle's motion to predict the evolution of the system and plan accordingly. To guarantee recursive feasibility of the closed-loop system, we computed a less conservative robust control invariant set with a numerical approach that computes the minimum required safe distance at each time step in accordance with the associated road grade data and the lead car states. The proposed controller is robust with respect to any aggressive braking of the lead vehicle as well as any arbitrary road slope.

We conducted simulations using realistic data of lead car's velocity and road grade profile. We verified through simulation for two application scenarios that the proposed controller improves the performance in terms of comfort, safety and energy efficiency compared to the baseline controller. In addition, we showed that our proposed ACC design is able to switch between CC and ACC mode automatically, smoothly and safely.

\section{Acknowledgement}
\label{secAcknowledgement}
The information, data, or work presented herein was funded in part by the Advanced Research Projects Agency-Energy (ARPA-E), U.S. Department of Energy, under Award Number DE-AR0000791. The views and opinions of authors expressed herein do not necessarily state or reflect those of the United States Government or any agency thereof.

\printbibliography

\end{document}